\documentclass[12pt]{article}
\usepackage{amsmath,amssymb,bm,graphicx}
\begin{document}

\begin{flushright}
\end{flushright}

\vskip 0.5 truecm

\begin{center}
{\Large{\bf Comment on the uncertainty relation with periodic boundary conditions}}
\end{center}
\vskip .5 truecm
\centerline{\bf  Kazuo Fujikawa }
\vskip .4 truecm
\centerline {\it Institute of Quantum Science, College of 
Science and Technology}
\centerline {\it Nihon University, Chiyoda-ku, Tokyo 101-8308, 
Japan}
\vskip 0.5 truecm


\begin{abstract}
The Kennard-type uncertainty relation $\Delta x\Delta p >\frac{\hbar}{2}$ is formulated for a free particle with given momentum $\langle \hat{p}\rangle$ inside a box with periodic boundary conditions in the large box limit. Our construction of a free particle state is analogous to that of the Bloch wave in a periodic potential. A simple Robertson-type relation, which minimizes the effect of the box boundary and may be useful in some practical applications, is also presented.
\end{abstract}


\section{Introduction}
It is well-known that the uncertainty relation~\cite{heisenberg, kennard, robertson} defined in a finite domain with a periodic (or anti-periodic) boundary condition needs to be treated with due care. For example, 
the Kennard-type relation\footnote{We call  the uncertainty relation defined in terms of standard deviations as the "Kennard-type relation" when the lower bound is state-independent~\cite{kennard}.} for the angular momentum 
$[L_{z}, \varphi]=\frac{\hbar}{i}$, which naively imlies $\Delta L_{z}\Delta \varphi\geq \hbar/2$, is easily violated since $\Delta L_{z}=0$ and $\Delta \varphi\leq 2\pi$ for the eigenstate of $L_{z}$. One can resolve this difficulty by a variety of ways~\cite{judge,judge2}, in particular, by noting that $e^{i\varphi}$ is periodic in $\varphi$ and unitary~\cite{louisell, susskind, carruthers}~\footnote{In contrast, a related issue of the hermitian photon phase operator, namely, the angular variable $\phi$ conjugate to the photon number operator $N=a^{\dagger}a$, which was the semi-classical basis of the quantization of the photon number to be an integer~\cite{dirac}, does not exist in principle, namely, one cannot define the unitary operator $e^{i\phi}$~\cite{louisell,susskind, carruthers, jackiw}. This is related to the notion of index associated with linear operators~\cite{fujikawa}.}.  

On the other hand, the periodic boundary condition inside a large box is used in many applications of quantum mechanics and also in many practical applications of field theory. Although the use of the periodic boundary condition in these cases is just for convenience, but the formulation of the Kennard-type uncertainty relation contains the same technical problem.  
The purpose of the present note is to define a free propagating particle in a box which naturally satisfies the conventional Kennard-type relation in the large box limit. 

If one uses a periodic boundary condition  $\psi(-\frac{L}{2},t)=\psi(\frac{L}{2},t)$ for a finite domain $-\frac{L}{2}\leq x \leq\frac{L}{2}$, the coordinate
does not send a periodic wave function to another  periodic function in general
\begin{eqnarray}
\psi(x,t)\rightarrow x\psi(x,t) 
\end{eqnarray}
since the simple coordinate variable $x$ does not preserve the periodic 
property. One may thus consider, for example,
\begin{eqnarray}
X(x)&=&x  \ \ \ {\rm for} \ \ \ -\frac{L}{2}\leq x <\frac{L}{2}\nonumber\\
    &=&-\frac{L}{2}
 \ \ \ {\rm for} \ \ \ x=\frac{L}{2}
\end{eqnarray}
and the periodic extension of this function to entire space $-\infty <x<\infty$ if one wishes~\cite{judge2, susskind}. In this case, one obtains
\begin{eqnarray}
i[p,X]=\hbar\left(1-L\delta(x-\frac{L}{2})\right)
\end{eqnarray}
in the interval $-\frac{L}{2}\leq x \leq \frac{L}{2}$.
This relation gives rise to
\begin{eqnarray}
\Delta p\Delta x\geq\frac{\hbar}{2}|1-L|\psi(\frac{L}{2},t)|^{2}|
\end{eqnarray}
if one performs the standard analysis in the  manner of Kennard~\cite{kennard} and Robertson~\cite{robertson} for a state vector $\psi(x,t)$. Note that we used $\Delta X=\Delta x$. This relation (4) is also obtained by the conventional analysis of the 
Kennard relation if one treats the surface term carefully
in the partial integration.

For the pure plane wave solution of the Schr\"{o}dinger equation of a free particle 
\begin{eqnarray}
\psi(x,t)=
\frac{1}{\sqrt{L}}\exp[i\frac{2\pi n}{L}x-i(\frac{2\pi n}{L})^{2}\hbar/(2m) t]
\end{eqnarray}
with an integer $n$, (4) gives an inequality 
\begin{eqnarray}
\Delta p\Delta x\geq 0
\end{eqnarray}
consistent with $\Delta p=0$ and $\Delta X=\Delta x=\sqrt{1/12}L$. However, the uncertainty relation one generally has in mind for    a free particle in the box normalization for 
large $L$ is not the above uncertainty relation (6).

For an explicit illustration, one may consider the wave packet 
\begin{eqnarray}
\psi_{n}(x)=\frac{1}{\sqrt{(1+2b^{2})L}}&\{&b\exp[i\frac{2\pi(n-1)}{L}x-i(\frac{2\pi(n-1)}{L})^{2}\hbar/(2m) t]\nonumber\\
&&+ \exp[i\frac{2\pi n}{L}x-i(\frac{2\pi n}{L})^{2}\hbar/(2m) t]\nonumber\\
&&+ b\exp[i\frac{2\pi(n+1)}{L}x-i(\frac{2\pi(n+1)}{L})^{2}\hbar/(2m) t]\}\nonumber\\
\end{eqnarray}
with a real parameter $b$, which is the solution of the Schr\"{o}dinger equation for a free particle
with mass $m$.
We then have  
\begin{eqnarray}
\Delta p =\sqrt{\frac{2b^{2}}{1+2b^{2}}}\frac{2\pi\hbar}{L}
\end{eqnarray}
which is independent of $t$ and $n$. On the other hand, obviously $\Delta x\sim L$
in the above wave packet although $\Delta x$ weakly depends on $t$ since the wave packet is oscillating in time. Namely, we have always the Kennard-type uncertainty relation
\begin{eqnarray}
\Delta p \Delta x\geq \frac{\hbar}{2}
\end{eqnarray}
if one understands the lower bound as a finite fraction of $\hbar$. 
This result is consistent with the relation (4) for $t=0$
\begin{eqnarray}
\Delta p \Delta x\geq \frac{\hbar}{2}|1-\frac{(1-2b)^{2}}{1+2b^{2}}|
=\frac{\hbar}{2}
\end{eqnarray}
if one uses $b=1/2$. However, for a suitably chosen time $t$, one can confirm that
the right-hand side of (4) can generally vanish. This fact prompted various authors to study a sensible definition of the Kennard-type relation  with periodic boundary conditions~\cite{judge,judge2,louisell,susskind, carruthers, jackiw}.

We re-examine this issue. Suppose that a state is given on the circle 
$S^{1}$ with a circumference $L$. Then one has a freedom in defining the coordinate on the circle. A sensible choice of the coordinate which minimizes the effects of the jump of the coordinate in (2), an artifact of introducing the box (or torus in general), may be  to choose the origin of the coordinate at the point where 
$|\psi(x,t)|^{2}$ is minimum. By cutting $S^{1}$ at such a point and taking the limit $L\rightarrow \rm large$, one may define a theory in an open space, which is the original object of our interest. One then has the Robertson-type relation~\footnote{We call the uncertainty relation whose lower bound explicitly depends on the used state as "Robertson-type" relation.} 
\begin{eqnarray}
\Delta p \Delta x\geq \frac{\hbar}{2}\left(1-L\times{\rm Min}_{x\in S^{1}}|\psi(x,t)|^{2}\right)
\end{eqnarray}
where the lower bound has a coordinate-independent meaning in the sense that it carries an intrinsic property of $\psi(x,t)$, and thus  state-dependent. The standard deviations $\Delta p$ and $\Delta x$ are evaluated in the interval $[x_{0}, x_{0}+L]$ where $x_{0}$ gives the minimum of $|\psi(x,t)|^{2}$ on $S^{1}$.  This relation imposes a self-consistency condition on the state defined on $S^{1}$ arising from the positive definite metric in the Hilbert space. Except for the pure plane wave (5), the lower bound in (11) is always away from 0.  Another convenient form of the relation (11) is 
\begin{eqnarray}
\Delta p \Delta x\geq \frac{\hbar}{2}\left(1-L\times{\rm Max}_{t}{\rm Min}_{x\in S^{1}}|\psi(x,t)|^{2}\right)
\end{eqnarray}
where ${\rm Max}_{t}{\rm Min}_{x\in S^{1}}|\psi(x,t)|^{2}$ stands for the maximum of  ${\rm Min}_{x\in S^{1}}|\psi(x,t)|^{2}$  with respect to the time $t$ and thus time-independent. 
In practice, one needs to know $|\psi(x,t)|^{2}$ with certain 
accuracy to evaluate $\langle x\rangle$ and $\langle x^{2}\rangle$ in the analysis of the uncertainty
relation. Thus the knowledge of $|\psi(x,t)|^{2}$ is a part of the analysis of the uncertainty relation and not an extra requirement. 

The relations (11) and (12) are the basis of our analysis. These  Robertson-type uncertainty relations are based on the algebra (3) by taking into account the special property of the circle $S^{1}$, and the lower bound is state-dependent.  The simple specification in (11) and (12) may be useful in some practical applications; for the 
generic wave packet such as (7), the lower bound in (12) is of the order of a finite fraction of $\hbar/2$ instead of precisely $\hbar/2$, which may be taken as the physical meaning of the uncertainty relation.

We here compare our approach with some representative formulations in the past. One of the basic formulations is to use the variables $\sin\frac{2\pi}{L}x$ and $\cos\frac{2\pi}{L}x$ in place of the ordinary coordinate $x$~\cite{louisell, susskind, carruthers}. One then has, for example,
\begin{eqnarray}
[p, (L/2)\sin\frac{2\pi}{L}x]=\frac{\hbar}{i}\pi \cos\frac{2\pi}{L}x
\end{eqnarray}
and thus 
\begin{eqnarray}
\Delta p\Delta\left((L/2)\sin\frac{2\pi}{L}x\right)\geq\frac{\hbar}{2}|\langle
\pi \cos\frac{2\pi}{L}x\rangle|.
\end{eqnarray}
This formulation is indispensable for the problem where no hermitian
angular variable $\phi$ (to be precise, no unitary $e^{i\phi}$) is defined but the hermitian analogues of 
$\cos\phi$ and $\sin\phi$ are defined such as in the photon phase operator~\cite{louisell, susskind, carruthers, jackiw, fujikawa}.  
But for our problem in the basically open space, this formulation is not quite convenient: For example, the above relation gives $\Delta(p)\Delta((L/2)\sin\frac{2\pi}{L}x)\geq 0$ while we actually have $\Delta p=2\pi\hbar/L$ and 
$\Delta\left((L/2)\sin\frac{2\pi}{L}x\right)=\sqrt{3}L/4$ for the state $\psi(x)=\sqrt{2/L}\sin\frac{2\pi}{L}x$. In contrast, our formulation gives
$\Delta p\Delta(x)\geq \hbar/2$ for the same state $\psi(x)=\sqrt{2/L}\sin\frac{2\pi}{L}x$.  

Another basic formulation initiated in~\cite{judge} is to define
\begin{eqnarray}
(\Delta x)_{\gamma}^{2}={\rm Min}_{\gamma\in S^{1}}\int_{-\frac{L}{2}}^{\frac{L}{2}}x^{2}|\psi(t,x+\gamma)|^{2}dx
\end{eqnarray}
which implies 
\begin{eqnarray}
\langle x\rangle&=&\int_{-\frac{L}{2}}^{\frac{L}{2}}x|\psi(t,x+\gamma)|^{2}dx\nonumber\\
&=&0
\end{eqnarray}
and 
\begin{eqnarray}
1-L|\psi(t,\frac{L}{2}+\gamma)|^{2}\geq 0
\end{eqnarray}
for such a value of $\gamma$; these two conditions are derived from $\frac{d}{d\gamma}(\Delta x)_{\gamma}^{2}=0$ and $\frac{d^{2}}{d\gamma^{2}}(\Delta x)_{\gamma}^{2}\geq 0$, respectively. The Robertson-like relation is then given by
\begin{eqnarray}
\Delta p(\Delta x)_{\gamma}\geq \frac{\hbar}{2}\left(1-L|\psi(t,\frac{L}{2}+\gamma)|^{2}\right)
\end{eqnarray}
and the lower bound vanishes only for the pure plane wave.
This formulation also exploits the coordinate independent property 
of the given state $\psi(t,x)$.

Both of (11) and (18) impose the consistency condition arising from the algebra (3) and the positive definite Hilbert space. From the point of view of a theory defined in the open space, these two definitions correspond to different ways of cutting the circle $S^{1}$ to define the open space. In the definition of (18) one cuts the circle $S^{1}$ such that the condition  $\langle x\rangle=0$ in (16) is satisfied for a given state $\psi(t,x)$, while (11) is defined by the cut of $S^{1}$ at the minimum of $|\psi(t,x)|$.  We have the relations
\begin{eqnarray}
\frac{\hbar}{2}\left(1-L\times{\rm Min}_{x\in S^{1}}|\psi(x,t)|^{2}\right)
\geq \frac{\hbar}{2}\left(1-L|\psi(t,\frac{L}{2}+\gamma)|^{2}\right)
\end{eqnarray}
and 
\begin{eqnarray}
\Delta p \Delta x\geq \Delta p(\Delta x)_{\gamma},
\end{eqnarray}
namely, (18) pays main attention to the actual minimum uncertainty product for the given state rather than the lower bound in the Robertson-type relation. The relation (18) is advantageous for an intrinsically periodic system such as the angular momentum.    

On the other hand, (11) pays more attention to the lower bound in the Robertson-type relation rather than the actual value of the uncertainty product. This is one important aspect of the uncertainty relation in the practical use. The relation (11) emphasizes how one can achieve the lower bound closest to the ordinary lower bound in the Kennard relation which is defined for normalizable states in the open space. The ultimate purpose of our use of the box normalization is to study the properties of general states in the open space in a mathematically  well-defined manner.

Coming back to the explicit example in (7), we illustrate the use of our formulation. We have
\begin{eqnarray}
|\psi_{n}(x,t)|^{2}=\frac{1}{(1+2b^{2})L}[(2b)^{2}\left(\cos 2\pi(\frac{x}{L}-n\alpha)+
\frac{\cos \pi\alpha}{2b}\right)^{2}+1-\cos^{2}\pi\alpha]
\end{eqnarray}
where
\begin{eqnarray}
\alpha=(\frac{2\pi}{L})^{2}\frac{2\hbar}{2m}t.
\end{eqnarray}
One thus has
\begin{eqnarray}
&&L\times{\rm Min}_{x\in S^{1}}|\psi(x,t)|^{2}\nonumber\\
&&=
\frac{1}{(1+2b^{2})}(1-\cos^{2}\pi\alpha) \ \ {\rm for} \ \ |\frac{\cos \pi\alpha}{2b}|\leq 1\nonumber\\
&&=\frac{1}{(1+2b^{2})}[(2b)^{2}\left(1-
|\frac{\cos \pi\alpha}{2b}|\right)^{2}+1-\cos^{2}\pi\alpha] \ \ {\rm for} \ \ \frac{1}{2b}\geq |\frac{\cos \pi\alpha}{2b}|\geq 1\nonumber\\
\end{eqnarray}
and the maximum of this quantity with respect to time is given by 
\begin{eqnarray}
{\rm Max }_{t}\{L\times{\rm Min}_{x\in S^{1}}|\psi(x,t)|^{2}\}=\frac{1}{1+2b^{2}}.
\end{eqnarray}
The final formula (12) for the wave packet in (7) is thus given by 
\begin{eqnarray}
\Delta p \Delta x\geq \frac{\hbar}{2}(1-\frac{1}{1+2b^{2}})
\end{eqnarray}
which is the Robertson-type relation for a particle  with  momentum $<p>=2\pi n\hbar/L$ and
$\Delta p =\sqrt{\frac{2b^{2}}{1+2b^{2}}}(2\pi\hbar/L)$ for
large $L$ and $n$ with fixed $<p>$. The relation (25) gives 
\begin{eqnarray}
\Delta p \Delta x\geq \frac{\hbar}{2}\times\frac{1}{3}
\end{eqnarray}
for $b=1/2$, for example; this relation needs to be satisfied for any $\Delta x$ calculated by our prescription independently of $t$. Physically, the assumption one makes is that the momentum resolution of the detector is larger than $\sim \pi\hbar/L$ for large enough $L$, which may suggest $b\geq 1/2$.

One may still notice that the mathematical limit $b=0$ in (25) gives rise to the 
relation (6) we rejected. As long as the basic algebraic relations in (3) and (13) are used, one generally ends in the Robertson-type relation where the lower bound is  state-dependent. Only exception is to use our prescription (11) combined with the zero in the wave function $|\psi(x,t)|$. Only for such a case, one can maintain the Kennard-type uncertainty relation where the lower bound is independent of the state vector. In this connection the basic idea of the large 
box normalization in field theory is instructive; we 
analyze physical properties which are "localized" relative to the large box itself which is introduced as a means to define the mathematics. For example, one may start with Green's  functions defined in a large box with periodic boundary conditions in interaction picture perturbation theory and then analyze their long distance behavior after first taking the limit $L\rightarrow {\rm large}$.  Likewise, we may avoid the direct use of the precise pure plane wave such as (5) in the analysis of the uncertainty relation by imposing certain locality requirement, although the idealized pure plane waves are useful and sufficient in many other physical applications.

\section{Bloch wave-like construction}

We here recall the basic motivation for using the box normalization in quantum mechanics. The standard formulation of quantum mechanics is based on  normalizable states in open space, but one important solution of the Schr\"{o}dinger equation, namely, the plane wave of the free Schr\"{o}dinger equation, among others, is not normalizable. One thus introduces an auxiliary box and defines the plane wave with the periodic boundary condition. This plane wave is also complete inside the box, namely, any periodic function inside the box is expanded in terms of plane waves. One may later take the large box limit to realize the open space. 

The plane wave which is the eigenstate of momentum operator causes the well-known complication in the analysis of the uncertainty relation as was discussed in the preceding section. 
The similar complication in the uncertainty relation also appears in the analysis of the angular momentum and phase operators in general. But the use of the box normalization in open space is of more technical character and thus one may deal with the complication in the box normalization by some technical means. 
Any free particle created in the laboratory is localized in space, and we 
take this property into account in the analysis of the Kennard-type 
relation.

We start with the solutions of the Schr\"{o}dinger equation for a free particle with the application of the periodic boundary condition (for $k\neq 0$)
\begin{eqnarray}
\psi_{n,k,\pm}(x,t)&=&\frac{1}{\sqrt{2L}}\{
\exp[i\frac{2\pi(n+k)}{L}x-i(\frac{2\pi(n+k)}{L})^{2}(\hbar/2m) t]\nonumber\\
&&\hspace{1cm}\pm\exp[i\frac{2\pi(n-k)}{L}x-i(\frac{2\pi(n-k)}{L})^{2}(\hbar/2m) t]\}\nonumber\\
&=&\frac{1}{\sqrt{2L}}\exp[i\frac{2\pi n}{L}x-i(\frac{2\pi}{L})^{2}(n^{2}+k^{2})(\hbar/2m) t]\nonumber\\
&&\hspace{1cm}\times\{\exp[ik\frac{2\pi}{L}x-ik(\frac{2\pi}{L})^{2}n(\hbar/m) t]
\nonumber\\
&&\hspace{1.2cm}\pm\exp[-ik\frac{2\pi}{L}x+ik(\frac{2\pi}{L})^{2}n(\hbar/m) t]\}.
\end{eqnarray}
If one sets $p_{n}=\frac{2\pi\hbar n}{L}$
this is written as 
\begin{eqnarray}
\psi_{n,k,+}(x,t)&=&\frac{2}{\sqrt{2L}}\exp[i\frac{p_{n}x}{\hbar}-i\frac{p_{n}^{2}+p_{k}^{2}}{2m\hbar} t]\nonumber\\
&&\hspace{1cm}\times\cos\left(k\frac{2\pi}{L}[x-(p_{n}/m) t]\right)\nonumber\\
\psi_{n,k,-}(x,t)&=&\frac{2i}{\sqrt{2L}}\exp[i\frac{p_{n}x}{\hbar}-i\frac{p_{n}^{2}+p_{k}^{2}}{2m\hbar} t]\nonumber\\
&&\hspace{1cm}\times\sin\left(k\frac{2\pi}{L}[x-(p_{n}/m) t]\right)
\end{eqnarray}
and 
\begin{eqnarray}
\psi_{n,0,+}(x,t)&=&\frac{1}{\sqrt{L}}\exp[i\frac{p_{n}x}{\hbar}-i\frac{p_{n}^{2}}{2m\hbar} t]
\end{eqnarray}
If one looks at these expressions at $t=0$, this construction is 
analogous to that of the Bloch wave with $p_{n}$ standing for the 
Bloch momentum and a complete set of periodic functions within the interval $[-\frac{L}{2},\frac{L}{2}]$. The difference is that the wave
packets in (28) are moving with the velocity $p_{n}/m$.

All the wave packets in (28) except for the pure plane wave in (29) satisfy
the Kennard-type uncertainty relation $\Delta p \Delta x\geq \frac{\hbar}{2}$ by our construction in (11). These relations correspond to the choice $b=\infty$ and $b=0$ in (25), respectively. 
We thus attempt to exclude the pure plane wave in (29) by replacing it by a more desirable state which still retains all the properties we expect for a free particle.

A way to avoid the pure plane wave is to retain only the second wave in (28), but now we consider that half of the box $0\leq x\leq\frac{L}{2}$ is our entire world
\begin{eqnarray}
\psi_{n,k,-}(x,t)&=&\frac{2i}{\sqrt{2(L/2)}}\exp[i\frac{p_{n}x}{\hbar}-i\frac{p_{n}^{2}+p_{k}^{2}}{2m\hbar} t]\nonumber\\
&&\hspace{1cm}\times\sin\left(k\frac{\pi}{L/2}[x-(p_{n}/m) t]\right).
\end{eqnarray}
Any function defined in the domain $0\leq x\leq\frac{L}{2}$ is formally extended to an odd function in the domain $-\frac{L}{2}\leq x\leq\frac{L}{2}$, and thus it is expanded in terms of 
$\frac{2i}{\sqrt{2L}}\sin(k\frac{2\pi}{L}x)$. This property is the basis of the construction in (30).

For a notational convenience, we re-write $L/2\rightarrow L$ in (30), and 
we obtain
\begin{eqnarray}
\psi_{n,k}(x,t)&\equiv&\frac{2i}{\sqrt{2L}}\exp[i\frac{p_{n}x}{\hbar}-i\frac{p_{n}^{2}+p_{k}^{2}}{2m\hbar} t]\nonumber\\
&&\hspace{1cm}\times\sin\left(k\frac{\pi}{L}[x-(p_{n}/m) t]\right)
\end{eqnarray}
defined in the interval $0\leq x\leq L$ with
\begin{eqnarray}
p_{n}\equiv\frac{\pi\hbar }{L}n
\end{eqnarray}
since the choice of the Bloch-like momentum is rather arbitrary. This construction when looked at $t=0$ is again analogous to the Bloch wave with a complete set of sine functions in the interval $0\leq x\leq L$ if one considers all positive integers $k$: To be precise, we have periodic waves in the basic interval $-L\leq x\leq L$ but we use only half of them defined in $0\leq x\leq L$. The construction, which is analogous to the Bloch wave, allows us to introduce the zero in the wave function to ensure the ordinary Kennard-like relation by our prescription (11) and at the same time to retain the notion of  momentum related to the plane wave.
From the point of view of the boundary condition, our wave-packets in (31) could  be constructed by waves with {\em periodic boundary conditions only} in the domain $[0, L]$ if one chooses both of $n$ and $k$ in (31) and (32) to be odd or even simultaneously, although a fixed common $n$ is more natural in analogy with the Bloch wave.

Practically this construction in (31) does not impose much restriction, since any localized (smooth) wave $\psi(x,t=0)$ in the sub-domain of $[0,L]$ can be extended to the wave in the entire domain $[0,L]$ by smoothly extending it to $\psi(x,t=0)=0$  at both ends of the domain $[0,L]$.  Such a wave $\psi(x,t=0)$ with an initial  momentum  
\begin{eqnarray}
\bar{p}_{n}=\int_{0}^{L} dx\frac{\hbar}{2i}[\psi(x,0)^{\dagger}\partial_{x}\psi(x,0)-(\partial_{x}\psi(x,0))^{\dagger}\psi(x,0)],
\end{eqnarray}
is expressed as a suitable superposition~\footnote{One may replace $p_{n}$ by a real number $\bar{p}_{n}$ in (31). In this case, one needs to use both periodic waves with even $k$ and anti-periodic waves with odd $k$ (or twisted boundary condition on the circle $S^{1}$) in the construction (31). The construction of wave packets then becomes very close to the Bloch wave. Note that the expansion coefficients in (34) are different from the coefficients in the expansion $\psi(x,t=0)=\sum_{k=1}\tilde{c}_{k}\frac{2i}{\sqrt{2L}}\sin( k\frac{\pi}{L}x)$. We use the expansion (34) since we want to emphasize a generalization of the plane wave. To preserve the plane wave behavior we need to satisfy $|\bar{p}_{n}| \gg |p_{k}|=k\frac{\pi}{L}$ for the main part of the expansion (34) in the large $L$ limit.} 
\begin{eqnarray}
\exp[-i\frac{\bar{p}_{n}x}{\hbar}]\psi(x,t=0)=\sum_{k=1}c_{k}\frac{2i}{\sqrt{2L}}\sin( k\frac{\pi}{L}x)
\end{eqnarray}
and the later time development is controlled by the Schr\"{o}dinger equation for a free particle
\begin{eqnarray}
\psi(x,t)=\sum_{k=1}c_{k}\psi_{\bar{n},k}(x,t)
\end{eqnarray}
where $\psi_{\bar{n},k}(x,t)$ stands for the wave packet in (31) with
$p_{n}$ replaced by $\bar{p}_{n}$.
When the time elapses, one may picture that such a localized wave  defined by (35) moves on the circle $\bar{S}^{1}$ for large $L$ with the velocity $\bar{p}_{n}/m$. Here $\bar{S}^{1}$ stands for the circle with a circumference $2L$ to account for the presence of periodic 
and anti-periodic waves, and for general $\bar{p}_{n}$
\begin{eqnarray}
\psi(x+2L,t)=\exp[i\frac{2\bar{p}_{n}L}{\hbar}]\psi(x,t)
\end{eqnarray}
which is the Bloch-like periodicity condition. Viewed in this manner, the part of the wave function $\psi(x,0)$ localized in $[-L,0]$ stands for the redundant freedom~\footnote{This is consistent since the probability flow at the points $x=(\bar{p}_{n}/m) t$ or $x=L+(\bar{p}_{n}/m) t$ up to a multiple of $2L$ is always zero.}.

Our solution (35) is written as 
\begin{eqnarray}
\psi(x,t)=\exp[i\frac{\bar{p}_{n}x}{\hbar}-i\frac{\bar{p}_{n}^{2}}{2m\hbar} t]\phi(x-(\bar{p}_{n}/m) t, t)
\end{eqnarray}
where $\phi(x,t)$ formally corresponds to a general solution of a free particle confined 
in a deep potential well with a width $0\leq x\leq L$,
\begin{eqnarray}
\phi(x,t)&=&\sum_{k=1}c_{k}\frac{2i}{\sqrt{2L}}\exp[-i\frac{p_{k}^{2}}{2m\hbar} t]\sin\left(\frac{k\pi}{L}x\right).
\end{eqnarray}
In our case, however, the deep potential well is moving with the velocity $(\bar{p}_{n}/m)$.
Not only each wave in (31) but also any superposition of the waves such as in (35) satisfy the condition
\begin{eqnarray}
{\rm Min}_{x\in \bar{S}^{1}}|\psi(x,t)|^{2}=0
\end{eqnarray}
for any $t$, namely at $x=(\bar{p}_{n}/m) t$ and $x=L+(\bar{p}_{n}/m) t$ up to a multiple of $2L$. This is the locality requirement in our formulation. One can thus define the Kennard-type relation
\begin{eqnarray}
\Delta p \Delta x\geq \frac{\hbar}{2}|1-L\times{\rm Min}_{x\in \bar{S}^{1}}|\psi(x,t)|^{2}|=\frac{\hbar}{2}
\end{eqnarray}
following our prescription in (11) where the integration domain to evaluate $\Delta p$ and $\Delta x$ is taken to be $[(\bar{p}_{n}/m) t, L + (\bar{p}_{n}/m) t]$. 

To be more explicit, we have 
\begin{eqnarray}
\langle p\rangle&=&\int_{(\bar{p}_{n}/m) t}^{L+(\bar{p}_{n}/m) t}\psi^{\star}(x,t)\frac{\hbar}{i}\partial_{x}\psi(x,t)dx\nonumber\\
&=&\bar{p}_{n}+\int_{0}^{L}\phi^{\star}(x,t)\frac{\hbar}{i}\partial_{x}\phi(x,t)dx\nonumber\\
&=&\bar{p}_{n}+\langle p\rangle_{\phi}(t)\nonumber\\
\langle p^{2}\rangle&=&\int_{(\bar{p}_{n}/m) t}^{L+(\bar{p}_{n}/m) t}\psi^{\star}(x,t)(\frac{\hbar}{i}\partial_{x})^{2}\psi(x,t)dx\nonumber\\
&=&\bar{p}_{n}^{2}+2\bar{p}_{n}\langle p\rangle_{\phi}(t)+\int_{0}^{L}\phi^{\star}(x,t)(\frac{\hbar}{i}\partial_{x})^{2}\phi(x,t)dx\nonumber\\
&=&\bar{p}_{n}^{2}+2\bar{p}_{n}\langle p\rangle_{\phi}(t)+\langle p^{2}\rangle_{\phi}(t)\nonumber\\
(\Delta p)^{2}&=&\langle p^{2}\rangle_{\phi}(t)-\left(\langle p\rangle_{\phi}(t)\right)^{2}
\end{eqnarray}
The average momentum $\langle p\rangle_{\phi}(t)$ for the state  
$\phi(x,t)$ is oscillating in time, and in our case $\langle p\rangle_{\phi}(0)=0$.
We analyze this issue in some detail; we have 
\begin{eqnarray}
\frac{d}{dt}\langle p\rangle_{\phi}(t)
&=&\int_{0}^{L}[\partial_{t}\phi^{\star}(x,t)\frac{\hbar}{i}\partial_{x}\phi(x,t)+\phi^{\star}(x,t)\frac{\hbar}{i}\partial_{x}\partial_{t}\phi(x,t)]dx\nonumber\\
&=&\frac{\hbar^{2}}{2m}\int_{0}^{L}[-\partial^{2}_{x}\phi^{\star}(x,t)\partial_{x}\phi(x,t)+\phi^{\star}(x,t)\partial_{x}\partial^{2}_{x}\phi(x,t)]dx\nonumber\\
&=&\frac{\hbar^{2}}{2m}[|\partial_{x}\phi(0,t)|^{2}-|\partial_{x}\phi(L,t)|^{2}]\nonumber\\
&=&\frac{1}{mL}\{|\sum_{k=1}c_{k}p_{k}\exp[-i\frac{p_{k}^{2}}{2m\hbar} t]|^{2}\nonumber\\
&&\hspace{1cm} -|\sum_{k=1}(-1)^{k}c_{k}p_{k}\exp[-i\frac{p_{k}^{2}}{2m\hbar} t]|^{2}\}.
\end{eqnarray}
If the condition  $|\partial_{x}\phi(0,t)|^{2}=|\partial_{x}\phi(L,t)|^{2}$ is satisfied $\langle p\rangle_{\phi}(t)$ becomes time independent, but this condition is not satisfied in general. Physically this means that the particle oscillates inside the deep potential. In our context, the edges of the box exert a force on the free particle inside due to the Dirichlet-type boundary condition. We now assume that $\phi(x,t)$ is {\em smooth} in the domain $[0,L]$ for any time $t$. This implies that both of $\partial_{x}\phi(0,t)$ and $\partial_{x}\phi(L,t)$ are finite, but for a technical reason we make a stronger assumption 
\begin{eqnarray}
\sum_{k=1}|c_{k}p_{k}|<\infty
\end{eqnarray}
which implies that the series expansion of  $\partial_{x}\phi(x,t)$ is absolutely convergent.
The right-hand side of (42) then goes to zero for large $L$. Namely, the boundary effect of the box normalization diminishes for a large box limit. (We expect that this property is satisfied by a condition weaker than (43).) Incidentally, $\int_{0}^{L}dx|\hbar\partial_{x}\phi(x,t)|^{2}=\sum_{k=1}|c_{k}p_{k}|^{2}<\infty$.

We also have 
\begin{eqnarray}
\langle x\rangle&=&\int_{(\bar{p}_{n}/m) t}^{L+(\bar{p}_{n}/m) t}x|\phi(x-(\bar{p}_{n}/m) t,t)|^{2}dx\nonumber\\
&=&\frac{\bar{p}_{n}}{m} t + \int_{0}^{L}x|\phi(x,t)|^{2}dx\nonumber\\
&=&\frac{\bar{p}_{n}}{m} t + \langle x\rangle_{\phi}(t),\nonumber
\end{eqnarray}
\begin{eqnarray}
\langle x^{2}\rangle&=&\int_{(\bar{p}_{n}/m) t}^{L+(\bar{p}_{n}/m) t}x^{2}|\phi(x-(\bar{p}_{n}/m) t,t)|^{2}dx\nonumber\\
&=&(\frac{\bar{p}_{n}}{m} t)^{2}+2(\frac{\bar{p}_{n}}{m} t)\langle x\rangle_{\phi}(t)+\langle x^{2}\rangle_{\phi}(t),\nonumber\\
(\Delta x)^{2}&=&\langle x^{2}\rangle_{\phi}(t)-(\langle x\rangle_{\phi}(t))^{2}.
\end{eqnarray}
Namely, the uncertainty product $\Delta x\Delta p$ for the state $\psi(x,t)$ is identical to that of the state $\phi(x,t)$. One can confirm Ehrenfest's relation in the sense  $\frac{d}{dt}\langle x\rangle=\langle p\rangle/m$.

Of course, one can expand the localized state $\psi(x,t)$ in terms of the pure plane waves in (5) or the set of states in (28) and (29), and still one can satisfy the Kennard-type relation for the state $\psi(x,t)$. The difference is that one can define the Kennard-type relation for all the component wave packets also in our construction. This property makes the analysis of the Kennard-type relation more reliable and practically useful. In many cases of the analysis of the uncertainty relation, one may want to retain a particle picture propagating in space and our construction provides such a picture. In the measurement process, one performs the measurement by a detector with a size much smaller than the box. One thus deals with  a transition from the above $\psi(x,t)$ or any of the basic waves in (31) to a smaller localized wave inside the box, and such a process is handled with the expansion (35) applied to the smaller localized wave by preserving the Kennard-type relation.   

For the elementary solution $\psi_{n,k}(x,t)$ in (31) we have
\begin{eqnarray}
\langle p\rangle&=&\int_{(p_{n}/m) t}^{L+(p_{n}/m) t}\psi^{\star}_{n,k}(x,t)\frac{\hbar}{i}\partial_{x}\psi_{n,k}(x,t)dx\nonumber\\
&=&p_{n}\nonumber\\
\langle p^{2}\rangle&=&\int_{(p_{n}/m) t}^{L+(p_{n}/m) t}\psi^{\star}_{n,k}(x,t)(\frac{\hbar}{i}\partial_{x})^{2}\psi_{n,k}(x,t)dx\nonumber\\
&=&\frac{1}{2}[(p_{n}+p_{k})^{2}+(p_{n}-p_{k})^{2}]\nonumber\\
\Delta p&=&\sqrt{p_{k}^{2}}=k\frac{\pi\hbar}{L}
\end{eqnarray}
and
\begin{eqnarray}
\langle x\rangle&=&\frac{2}{L}\int_{(p_{n}/m) t}^{L+(p_{n}/m) t}x\sin^{2}\left(k\frac{\pi}{L}[x-(p_{n}/m) t]\right)dx\nonumber\\
&=&\frac{L}{2}+\frac{p_{n}}{m} t,\nonumber\\
\langle x^{2}\rangle&=&\frac{2}{L}\int_{(p_{n}/m) t}^{L+(p_{n}/m) t}x^{2}\sin^{2}\left(k\frac{\pi}{L}[x-(p_{n}/m) t]\right)dx\nonumber\\
&=&\frac{L^{2}}{3}-\frac{2L^{2}}{(2\pi k)^{2}}+2(\frac{L}{2})(\frac{p_{n}}{m} t) +(\frac{p_{n}}{m} t)^{2}, \nonumber\\
\Delta x&=&\frac{L}{2\sqrt{3}}\sqrt{1-\frac{24}{(2\pi k)^{2}}}.
\end{eqnarray}
Thus
\begin{eqnarray}
\Delta x\Delta p=\frac{\pi\hbar}{\sqrt{3}}\sqrt{k^{2}-\frac{24}{(2\pi )^{2}}}.
\end{eqnarray}
The choice $k=1$ gives the minimum uncertainty state in our construction, and we have
\begin{eqnarray}
\Delta x\Delta p=\frac{\pi\hbar}{2\sqrt{3}}\sqrt{1-\frac{24}{(2\pi )^{2}}}\ >\frac{\hbar}{2}.
\end{eqnarray}
The numerical value of the uncertainty product 
$\Delta x\Delta p$ in (48) is close to the lower bound $\frac{\hbar}{2}$.
Our Kennard-type uncertainty relation gives rise to the prediction
which one normally expects in the applications of the box normalization with periodic boundary conditions for a free particle,  namely,
\begin{eqnarray}
\Delta p\sim \pi\hbar/L \ \ {\rm and} \ \   \Delta x \sim L 
\end{eqnarray}
and the well-defined notion of momentum $\langle p\rangle=p_{n}$ in the large $L$ limit.
The large $L$ (and large $n$) limit of our wave packet (31) then corresponds to the "physical" plane wave or a free particle in the conventional sense, in place of the mathematical pure plane wave in (5). The simplest candidate for a free particle with the momentum 
$\langle p\rangle=p_{n}$ in the box with the minimum uncertainty product in our construction is
\begin{eqnarray}
\psi_{n,1}(x,t)
&=&\frac{2i}{\sqrt{2L}}\exp[i\frac{p_{n}x}{\hbar}-i\frac{p_{n}^{2}+p_{1}^{2}}{2m\hbar} t]\nonumber\\
&&\hspace{1cm}\times\sin\left(\frac{\pi}{L}[x-(p_{n}/m) t]\right).
\end{eqnarray}
The behavior of this wave in the finite neighborhood of any fixed point which is far from both ends of the box, such as $x=\frac{L}{2}+\delta x$,  for large $L$ ( and large $n$) is close to the pure plane wave (5). This is what we expect for an actual free particle defined in a large box which was introduced as a convenient means.
 
\section{Conclusion}
One generally presumes that main physical properties are insensitive to 
the precise boundary condition (or other artifacts of the box normalization) for a large enough box except for the topological property such as the winding number~\footnote{The pure plane wave (5) may be understood as related to the winding number of the mapping $S^{1}\rightarrow U(1)$. 
}. It is interesting that the Kennard-type uncertainty relation is sensitive to the precise boundary condition. The periodic (or anti-periodic) boundary condition generally leads to the Robertson-type uncertainty relation such as (14) and (18), but our use of the box normalization is of  technical character and thus we sought for a technical way to preserve the Kennard-type relation inside a box. We exploited the fact that any free particle created in the laboratory is localized in space.
Our analysis of the Kennard-type relation shows how to construct a physical free particle state in the box normalization with periodic boundary conditions for large $L$.  Our construction of a free particle state is not unique, but it is based on a complete set of wave packets and thus it is as general as the use of  pure plane waves for a free particle for which, however, the construction of the Kennard-type relation has a  technical problem.  In some practical applications, our simple specification of the Robertson-type relation in (11) and (12) with the state-dependent lower bound may also be useful.

\end{document}